\documentclass[11pt]{article}
\usepackage{color}
\usepackage[colorlinks, linkcolor=color1, anchorcolor=blue, citecolor=color1]{hyperref}
\usepackage{geometry}
\usepackage{sectsty}
\sectionfont{\large}
\subsectionfont{\normalsize}

\usepackage{hyperref}

\usepackage{amssymb}
\usepackage{amsmath}
\usepackage[lined,boxed,commentsnumbered, ruled]{algorithm2e}
\usepackage{mathrsfs}
\usepackage{algorithmic}
\usepackage{bm}

\usepackage{pgfplots}
\usepackage{tikz}
\usetikzlibrary{arrows}
\usepackage{subfigure}
\usepackage{graphicx,booktabs,multirow}

\definecolor{colorhkust}{RGB}{20,43,140}
\definecolor{colortsinghua}{RGB}{116,52,129}
\definecolor{color1}{RGB}{128,0,0}

\newcommand{\rev}{\color{black}}

\date{}

\begin{document}

\title{{Generalized Sparse and Low-Rank Optimization for Ultra-Dense Networks}}
\author{Yuanming~Shi, Jun Zhang, Wei Chen, and Khaled B. Letaief}
{\let\thefootnote\relax\footnotetext{{This work was supported in part by the National
Nature Science Foundation of China (NSFC) under Grant No. 61601290, Shanghai Sailing Program under Grant No. 16YF1407700, the Hong Kong Research Grant Council under Grant No. 16200214, the National Natural Science Foundation of China under Project Nos. 61671269 and 61621091, the Chinese National 973 Program under Project No. 2013CB336600, and the 10000-Talent Program of China.}}}

{\let\thefootnote\relax\footnotetext{{Y. Shi is with the School of Information Science and Technology,
ShanghaiTech University, Shanghai, China (e-mail: shiym@shanghaitech.edu.cn).} }}

{\let\thefootnote\relax\footnotetext{{J. Zhang is with the Department of Electronic and Computer
Engineering, Hong Kong University of Science and Technology, Hong Kong (e-mail:
eejzhang@ust.hk).}}}

{\let\thefootnote\relax\footnotetext{{W. Chen is with the Department of Electronic
Engineering, Tsinghua University, Beijing, China (e-mail: wchen@tsinghua.edu.cn).}}}

{\let\thefootnote\relax\footnotetext{{K. B. Letaief is with Hamad bin Khalifa University,
Ar-Rayyan, Qatar, and also with the Hong Kong University of Science and Technology,
Hong Kong (e-mail: kletaief@hbku.edu.qa; eekhaled@ust.hk).}}}

\maketitle


\maketitle

\begin{abstract}
Ultra-dense network (UDN) is a promising technology to further evolve wireless networks and meet the diverse performance requirements of 5G networks. With abundant access points, each with communication, computation and storage resources, UDN brings unprecedented benefits, including significant improvement in  network spectral efficiency and energy efficiency, greatly reduced latency to enable novel mobile applications, and the capability of providing massive access for Internet of Things (IoT) devices. However, such great promises come with formidable research challenges. To design and operate such complex networks with various types of resources, efficient and innovative methodologies will be needed. This motivates the recent introduction of highly structured and generalizable models for network optimization.
In this article, we present some recently proposed  large-scale sparse and low-rank frameworks for optimizing UDNs, supported by various motivating applications. A special attention is paid on algorithmic approaches to deal with nonconvex objective functions and constraints, as well as computational scalability.
\end{abstract}


\section{Introduction}

As mobile data traffic keeps growing at an exponential rate, and mobile applications pose more and more stringent and diverse requirements, wireless networks are facing unprecedented pressures. To further evolve wireless networks and maintain their competitiveness, network infrastructure densification stands out as a promising approach. By deploying more radio access points, supplemented with storage and computational capabilities, we can not only increase network capacity, but also improve network energy efficiency, enable low-latency mobile applications, and provide access for massive mobile devices. Such ultra-dense network (UDN) provides an ideal platform to develop disruptive proposals to advance wireless information technologies, including cloud radio access networks (C-RANs), wireless edge caching, and mobile edge computing. These are achieved by leveraging innovative ideas in different areas, such as software-defined networking, network function virtualization, content-centric networking, cloud and fog computing.

By enabling capabilities of  cloud computing and software-defined networking, UDNs can easily support {\it{C-RAN}}  as an effective network architecture to exploit the benefits of  network densification via centralized signal processing and interference management \cite{Tony_CRAN17, Haijun_WCmag16}. This is achieved by moving the baseband  processing functionality to the cloud data center via high-capacity fronthaul links, supported by massively deployed low-cost remote radio heads (RRHs). Meanwhile, the Internet is shifting
from the ``connection-centric" mode to the ``content-centric" mode to support high-volume content
delivery \cite{Ververidis_CST14}. By enabling content caching at radio access points, i.e., {\it{wireless edge caching}}, UDNs can assist the Internet architecture evolution and achieve more efficient content delivery for mobile users \cite{Debbah_CMag14}. Another trend is the  increasing computation intensity in mobile applications, which puts a heavy burden on resource-constrained mobile devices. {\it{Mobile edge computing}} was recently proposed as a promising solution, by offloading computation tasks of mobile applications to servers at nearby access points. It avoids excessive propagation delay in the backbone network, compared to mobile cloud computing, and thus enables latency-critical applications. All of these systems are built upon the UDN platform, which enables integration of the storage, computing, control and networking
functionalities at the ubiquitous access points. {\rev{In particular, C-RANs serve the purpose of providing higher data rates,
while mobile edge caching and computing networks enable low-latency content delivery and mobile applications.}}

\begin{table}[t]\footnotesize
\renewcommand{\arraystretch}{1.3}
\caption{Generalized Sparse and Low-Rank Optimization for UDNs}
\label{table_srank}
\centering
\begin{tabular}{l|l|l}
Models & Structured Sparse Optimization (\ref{sparse}) & Generalized Low-Rank Optimization (\ref{lowrank})\\
\hline
\multirow{4}{*}{Applications} & Large-scale network adaptation:
& Network optimization with side information:
 \\
&1. Network power minimization  & 1. Topological interference management
\\
&2. User admission control & 2. Wireless distributed computing\\
& 3. Active user detection & 3. Mobile edge
caching \\
\hline
\multirow{3}{*}{Algorithms} & Convex optimization solver \cite{Yuanming_LargeSOCP2014}: 
& Riemannian optimization solver \cite{manopt}: 
 \\
&1. $\mathcal{O}(1/k)$ convergence rate: ($k$: $\#$ {\textrm{iterations}})  & 1. Superlinear convergence rate with conjugate gradient
\\
&2. Subspace projection per iteration & 2. Quadratic convergence rate with trust-region\\
& 3. Parallel cone projection per iteration & 3. Compute Riemannian gradient and Hessian per iteration

\end{tabular}
\label{parameter1}
\vspace*{-15pt}
\end{table}

However, all the  emerging
networking paradigms associated with UDNs bring formidable challenges to network optimization, signal processing and resource allocation, given the highly complex network topology, the massive amount of required side information, and the high computational requirement. Typical design problems are nonconvex in nature, and of enormously large scales, i.e., with large numbers of constraints and optimization variables. For examples, the uncertainty or estimation error in the available channel state information (CSI) yields nonconvex quality-of-service (QoS) constraints, while such network performance metrics as sum throughput and energy efficiency lead to nonconvex objective functions. Thus effective and scalable design methodologies, with the capability of handling nonconvex constraints and objectives, will be needed to fully exploit the benefits of UDNs. The aim of this article is to present recent advances in sparse and low-rank techniques for optimizing dense wireless networks \cite{Yuanming_TWC2014, Yuanming_2016LRMCTWC, Thomas_IEEEAccess20155G}, with a comprehensive coverage including modeling, algorithm design, and theoretical analysis. We identify two representative classes of design problems in UDNs, i.e., {large-scale network adaption} and {side information assisted network optimization}.  

The first class of design problems are for the efficient network adaptation in UDNs, including  radio access point selection
\cite{Yuanming_TWC2014}, backhaul data assignment, user
admission control, user association \cite{Haijun_JSAC17}, and active user detection \cite{Thomas_IEEEAccess20155G}. Such large-scale network adaptation problems involve both discrete
and continuous decision variables, which motivates us to enforce sparsity
structures in the solutions. The success of the structured sparse optimization for network adaptation comes from the key observation that such adaptation can be achieved by enforcing structured sparsities in the solution, which will be presented  in Section {\ref{sparseopt}} in details. The second class of design problems involve how to effectively utilize the available side information for network optimization, including topological interference management \cite{Jafar_TIT2013TIM}, wireless
distributed computing \cite{Ali_ComMag17}, and mobile edge caching \cite{Debbah_CMag14}.
Network side information is critical to design UDNs, and it can take various forms, such as the network connectivity information, cache content placement at access points, and locally computed  intermediate values in wireless distributed computing. In Section {\ref{lowrank1}}, we will present a general incomplete matrix framework to model
various network side information, which leads to a unified network performance metric via the rank of the modeling matrix for optimizing UDNs.

Although the structured sparse and low-rank techniques enjoy the benefits of modeling flexibility, the sparse function and rank function are nonconvex, which brings computational challenges \cite{Wainwright2014structured, Romberg_JSTSP16lrmc}. Furthermore, typical optimization problems in UDNs bear complicated structures,
which make most of the existing algorithms and theoretical results inapplicable. To address these algorithmic challenges, we present various convexification procedures for both objectives and constraints throughout our discussion. Moreover, scalable convex optimization algorithms and nonconvex optimization techniques, such as Riemannian optimization, will be presented in Section {\ref{alg}}. This
article shall serve the purpose of providing network modeling methodologies
and scalable computational tools for optimizing complex UDNs, {\rev{as summarized in Table {\ref{table_srank}}}}.

\section{Structured Sparse Optimization for Large-Scale Network Adaptation}
\label{sparseopt}
In UDNs, to effectively utilize densely deployed access points \ to support massive mobile devices,
large-scale network adaptation will play a pivotal role. For various network adaptation problems in UDNs,
the solution vector is expected to be sparse in a structured manner, e.g., radio access point selection results in a group sparsity structure. To illustrate the powerfulness of the generalized sparse representation and scalable optimization  paradigms, in this section, we present representative examples of group sparse beamforming for green C-RANs, and structured sparse optimization for active users detection and user admission control.

\subsection{Generalized Structured Sparse Models}
\begin{figure}[t]
\center
\includegraphics[scale = 0.6]{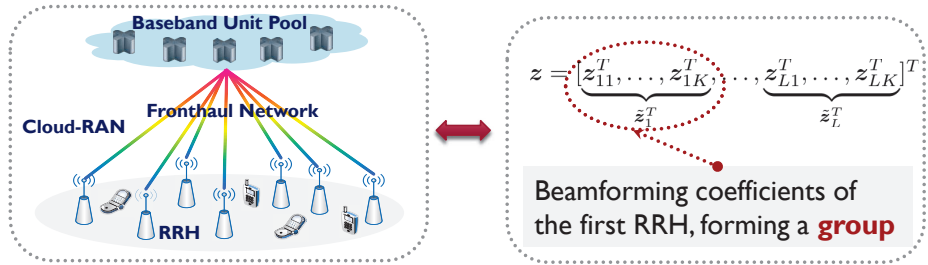}
\caption{Group sparse beamforming for green Cloud-RAN design, with $L$ RRHs and $K$ users. Switching off the $l$-th RRH and the corresponding fronthaul link corresponds to setting all the beamforming coefficients in the $l$-th beamforming coefficients group $\tilde{\bm{z}}_l=[\bm{z}_{l1},\dots,\bm{z}_{lK}]$ to be zeros simultaneously, where $\bm{z}_{lk}$ is the transmit beamforming vector from RRH $l$ to mobile user $k$. {\rev{Note that, when switching off one RRH, the remaining RRHs need to support the QoS requirements for all the mobile users.}}}
\label{cran}
\end{figure}
In this part, two motivating applications of generalized sparse models for large-scale network adaptation are presented. 

\subsubsection{Large-Scale Structured Optimization}
\label{lssp}
We take green C-RAN as an example to  illustrate structured optimization for network adaptation. In C-RANs, the network power consumption
consists of the transmit power of active RRHs and the power of the corresponding
active fronthaul links. By exploiting the spatial and temporal data traffic fluctuation, network adaption via dynamically switching off RRHs and the associated
fronthaul links can significantly reduce the network power consumption. To minimize the network power of a C-RAN, we need to optimize over both the discrete variables (i.e., the selection
of RRHs and fronthaul links) and continuous variables (i.e., downlink beamforming coefficients), yielding a mixed combinatorial optimization problems, which is highly-intractable. To support
efficient algorithm design and analysis, a principled group sparse beamforming
framework was proposed in \cite{Yuanming_TWC2014} by enforcing the group
sparsity structure in the solution vectors. This is achieved by a group sparsity representation of
the discrete optimization variables for RRHs selection as shown in Fig. {\ref{cran}}. Specifically, by regarding all the beamforming
coefficients of one RRH as a group, switching off this RRH corresponds to setting all the associated beamforming
coefficients in the same group to be zero simultaneously. We thus enforce the group sparsity structure in the aggregative beamforming vector to guide switching off the corresponding RRHs to minimize the network power consumption. Similar to group sparse beamforming for RRH selection, there is a corresponding user side node selection problem. With  crowded mobile devices, it is critical to maximize the user capacity, i.e., the number
of admitted users. This {\emph{user admission}} problem
is equivalent to minimizing the number of violated QoS  constraints ({\rev{modeled as $g_i(\bm{x})\le 0$ for the $i$-th user and may be infeasible}}), which can further be modeled as minimizing the individual sparsity
of the auxiliary vector {\rev{$\bm{z}=[z_i]$ with $z_i\ge 0$}} indicating the violations of the QoS constraints. {\rev{That is, for the constraint $g_i(\bm{x})\le z_i$ (always feasible as the auxiliary variable $z_i\ge 0$), $z_i=0$ indicates that the original QoS constraint $g_i(\bm{x})\le 0$ is feasible, while $z_i>0$ indicates that the original QoS constraint $g_i(\bm{x})\le 0$ is infeasible}}. Therefore, by enforcing this structured sparsity in the solution, user admission can be effectively handled.

\subsubsection{High-Dimensional Structured Estimation}
\label{activeuser}
\begin{figure}[t]
\begin{tabular}{cc}
\noindent \begin{minipage}[b]{0.4\hsize}
\centering
\includegraphics[width=\hsize]{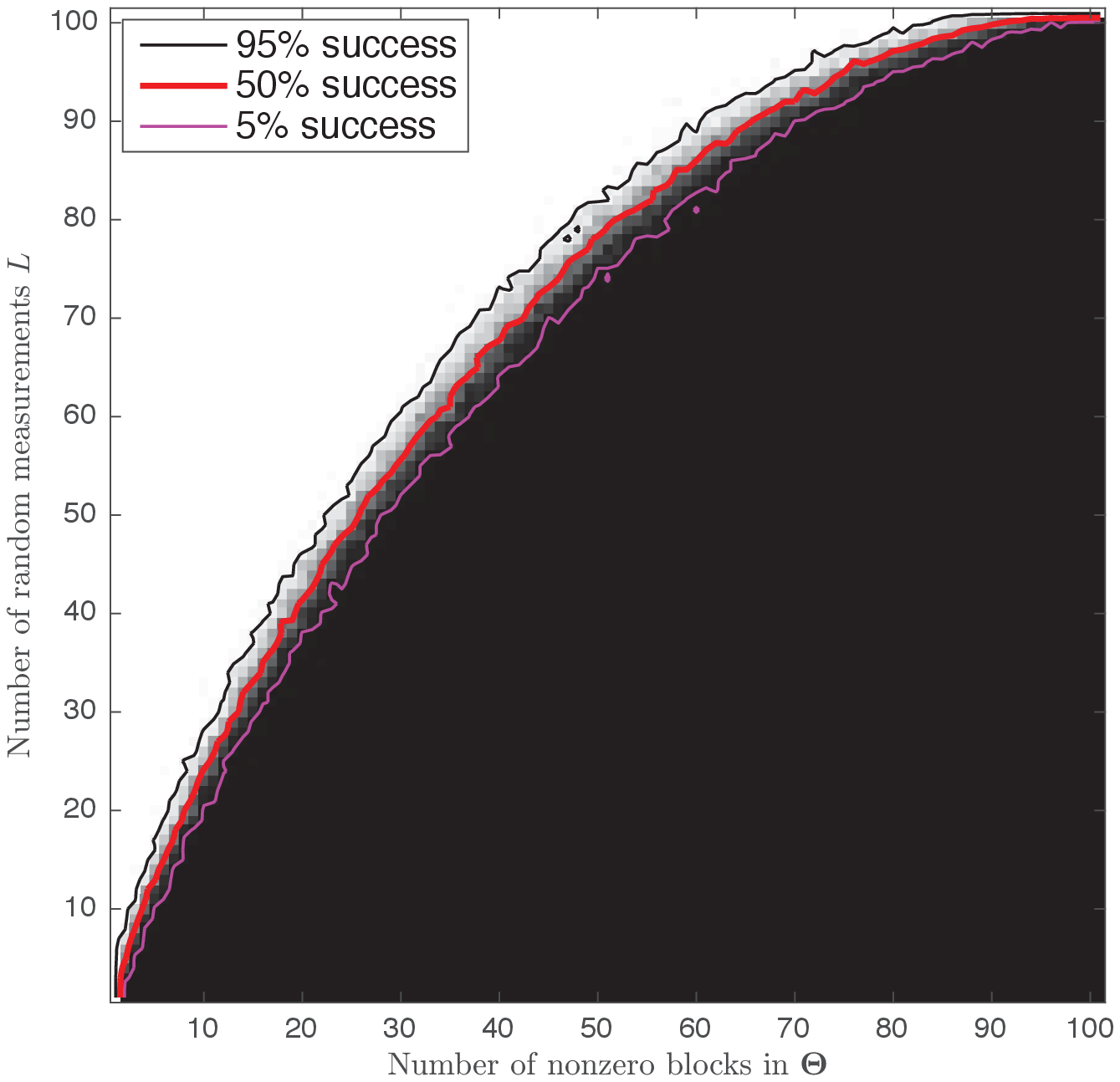}
{\scriptsize (a) Phase transitions in noiseless scenario.}
 \end{minipage}
 \hspace*{0.5cm}
 \noindent \begin{minipage}[b]{0.49\hsize}
\centering
\includegraphics[width=\hsize]{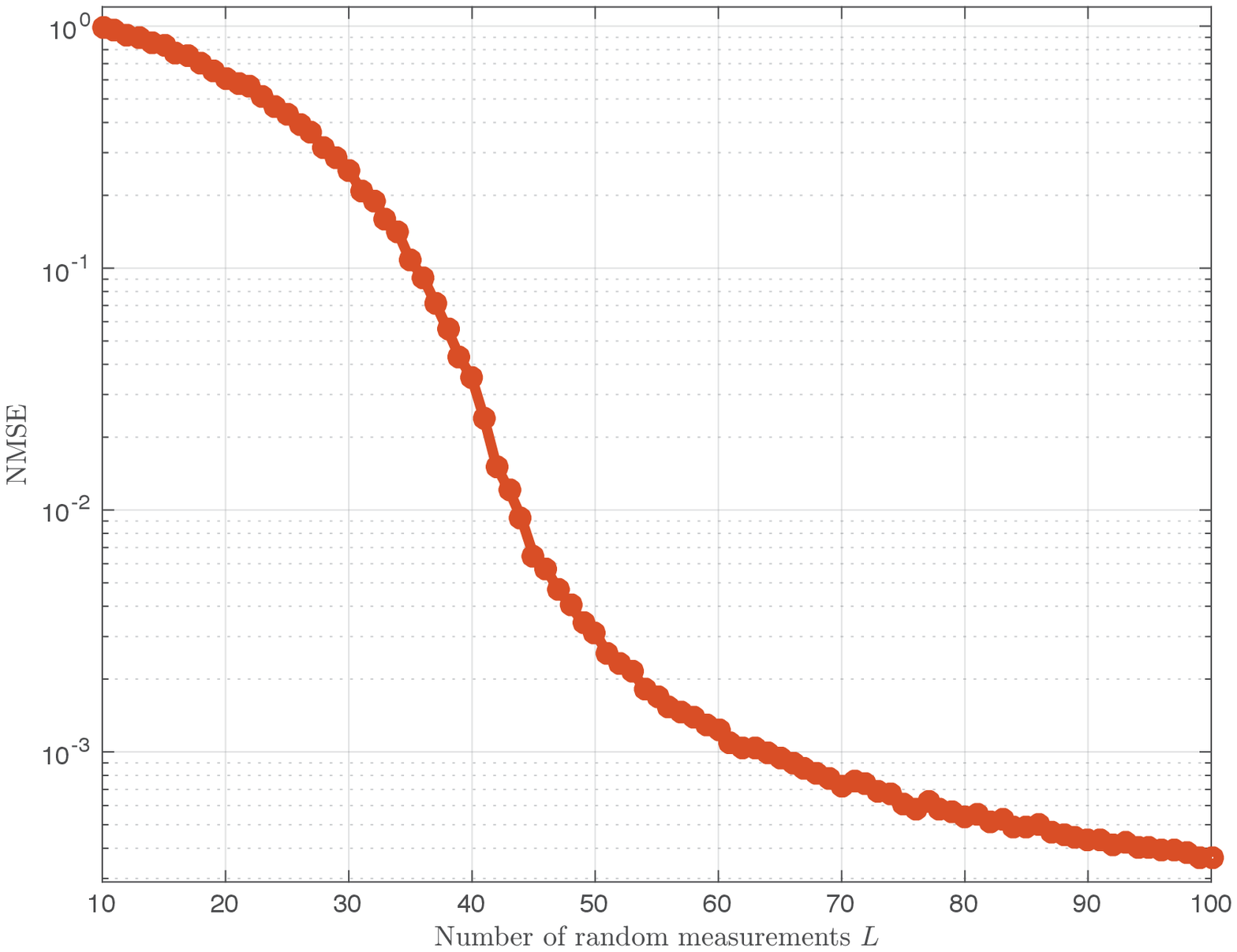}
{\scriptsize (b) NMSE in noisy scenario.}
 \end{minipage}
 \end{tabular}
\caption{(a) Phase transitions for structured sparse estimation
for massive device connectivity, given the random measurement $\bm{Y}=\bm{\Theta
Q}$ with $N=100$ and $M=2$. The heat map indicates the empirical probability of success (black=0$\%$; white=100$\%$). (b) NMSE of estimating $\bm{\Theta}$, given the noisy model $\bm{Y}=\bm{\Theta Q}+\bm{W}$ with $N=100$, $M=2$, $K=20$ and $W_{ij}\sim\mathcal{CN}(0,0.01)$. Each entry in  $\bm{Q}\in\mathbb{C}^{N\times L}$ is distributed as $Q_{ij}\sim\mathcal{CN}(0,1)$ with $L$ as the number of random measurements.}
\label{massivedevice}
\end{figure}

With limited radio resources, it is challenging to support massive device connectivity  for such applications as IoT. Fortunately,  only part of the massive devices will be active at a time given the sporadic traffic for the emerging applications (e.g., machine-type communications, Internet-of-Things (IoT)) \cite{Thomas_IEEEAccess20155G}. Active user detection is thus a key problem for providing massive connectivity in UDNs, which turns out to be a structured sparse estimation problem. Specifically, suppose we have $N$ single-antenna mobile devices ($K$ of which are active) and one $M$-antenna base station (BS). The received signal  at the BS has the form $\bm{Y}=\bm{H}\bm{\Sigma}\bm{Q}+\bm{W}$, where $\bm{\Sigma}\in\mathbb{R}^{N\times N}$ is the unknown diagonal activity matrix with $K$ non-zero diagonals whose positions are to be estimated, $\bm{H}\in\mathbb{C}^{M\times N}$ is the unknown channel matrix from all the devices to the BS, $\bm{Q}\in\mathbb{C}^{N\times L}$ is the known pilot matrix with training length $L$, and $\bm{W}$ is the additive noise. We thus need to simultaneously estimate the channel matrix $\bm{H}$ and $\bm{\Sigma}$, which poses a great challenge. We observe that detecting the active users is equivalent to estimating the group sparsity structure of the combined matrix $\bm{\Theta}=\bm{H\Sigma}\in\mathbb{C}^{M\times N}$, which has a group structured sparsity in columns of matrix $\bm{\Theta}$, induced by the structure of $\bm{\Sigma}$. That is, when mobile device $n$ is inactive, all the entries in the $n$-th column in matrix $\bm{\Theta}$ become zeros simultaneously. Due to the  limited radio resources, the training length $L$ will be much smaller than the channel dimension $N$, and thus, the estimation problem is ill-posed and yields a high-dimensional structured estimation problem. 

Fortunately, the embedded low-dimensional structure (i.e., the structured sparsity) can be algorithmically exploited to ensure the success for the high-dimensional structured estimation, as illustrated in Fig. {\ref{massivedevice}} for the behaviors of phase transitions and normalized mean square error
(NMSE) . Phase transition
defines a sharp change in the
behavior of a computational problem as its parameters vary. Convex
geometry and conic integral geometry provide principled ways to theoretical
predicate the phase transitions precisely \cite{Tropp_livingedge2014}. In particular, the phase transition phenomenon in Fig. {\ref{massivedevice}} (a) reveals the fundamental limits of sparsity recovery  in the best cases, i.e., without noise. Specifically, such study reveals that the required training length, or the number of measurements, depends on the sparsity level of $\bm{\Theta}$, and highly accurate user activity detection can be achieved with sufficient measurements. Fig. {\ref{massivedevice}} (b) further demonstrates that the low-dimensional structure can be exploited to significantly reduce the training length for active users detection even in the noisy scenarios.

\subsection{A Generalized Sparse Optimization Paradigm}

We have demonstrated that effective network adaptation can be achieved
by either inducing vector sparsity in the structured manner or estimating
the structured sparsity pattern. In this part, we provide  a generalized sparse optimization framework to algorithmically exploit the low-dimensional structures in UDNs. This is achieved by optimizing a constrained composite combinatorial objective:
\setlength\arraycolsep{2pt}
\begin{eqnarray}
\label{sparse}
\mathop {\rm{minimize}}_{\bm{z}\in\mathbb{C}^{n}}~~f(\bm{z}):=f_1({\rm{Supp}}(\bm{z}))+ f_2(\bm{z})~~~~{\rm{subject~to}}~~\bm{z}\in\mathcal{C},
\end{eqnarray}
where ${\rm{Supp}}(\bm{z})$ is the index set of non-zero coefficients of a vector $\bm{z}$, $f_1$ is a combinatorial positive-valued set-function to control the structured sparsity in $\bm{z}$, $f_2$ is a continuous convex function in $\bm{z}$ to represent the system performance such as transmit power consumption, and the constraint set $\mathcal{C}$ serves the purpose of modeling system constraints, e.g, transmit power constraints and QoS constraints. The most natural convex surrogate for a nonconvex function $f$ is its convex envelope, i.e., its tightest convex lower bound. The main motivation for convexifying function $f$ is that the convexified optimization problems make it possible to use the convex geometry theory \cite{Tropp_livingedge2014} to reveal benign properties about the globally optimal solutions, which can be computed with efficient algorithms. For example, the individual sparsity function with $\ell_0$-norm in $\bm{z}$ can be convexified to the $\ell_1$-norm. The group sparsity function can be convexified by the mixed $\ell_1/\ell_2$-norm.  More general convex relaxation results can be derived based on the principles of  convex analysis \cite{Tropp_livingedge2014}.  {\rev{Note that, it is critical to establish the optimality for various convex relaxation approaches in UDNs. For  example, for the nonconvex active user detection problem in Section \ref{activeuser}, the optimality condition can be established via the conic geometry approach in \cite{Tropp_livingedge2014}.}}

The constraint set $\mathcal{C}$ serves the purpose of modeling various QoS constraints including unicast beamforming, multicast beamforming, and stochastic beamforming, just to name a few. For example, the nonconvex QoS constraints for unicast beamforming can be equivalently transformed into convex second-order cone constraints \cite{Yuanming_TWC2014}. Furthermore, physical layer integration techniques can effectively improve the network performance via providing multicast services, which, however, yield nonconvex quadratic QoS constraints. The  semidefinite relaxation (SDR)  technique turns out to be effective to convexify the nonconvex quadratic constraints via lifting the original vector problem to higher matrix dimensions, followed by dropping the rank-one constraints. For stochastic beamforming with probabilistic QoS constraints due to CSI uncertainty, the probabilistic QoS constraints can be convexified based on the principles of the majorization-minimization procedure, yielding sequential convex approximations. In summary, the general formulation in (\ref{sparse}) enables efficient algorithm design and analysis for network adaptation in UDNs.

\section{Generalized Low-Rank Optimization with Network Side Information}
\label{lowrank1}
UDNs are highly complex to optimize, for which it is critical to exploit the available network side information. For example, network connectivity information,  cached content at the access points, and locally computed intermediate values, all serve as exploitable side information for efficiently designing coding and decoding in UDNs.
In this section, we provide a generalized low-rank matrix modeling framework to exploit the network side information, which helps to efficiently optimize across the communication, computation, and storage resources. To demonstrate the powerfulness of this framework, we present topological interference alignment as a concrete example and then extend it to  cache-aided interference channels and wireless distributed computing systems. A general low-rank optimization problem is then formulated by incorporating the network side information.

\subsection{Network Side Information Modeling via Incomplete Matrix}
\begin{figure}[t]
\begin{tabular}{cc}
\hspace*{0.5cm}
\noindent \begin{minipage}[b]{0.26\hsize}
\centering
\includegraphics[width=\hsize]{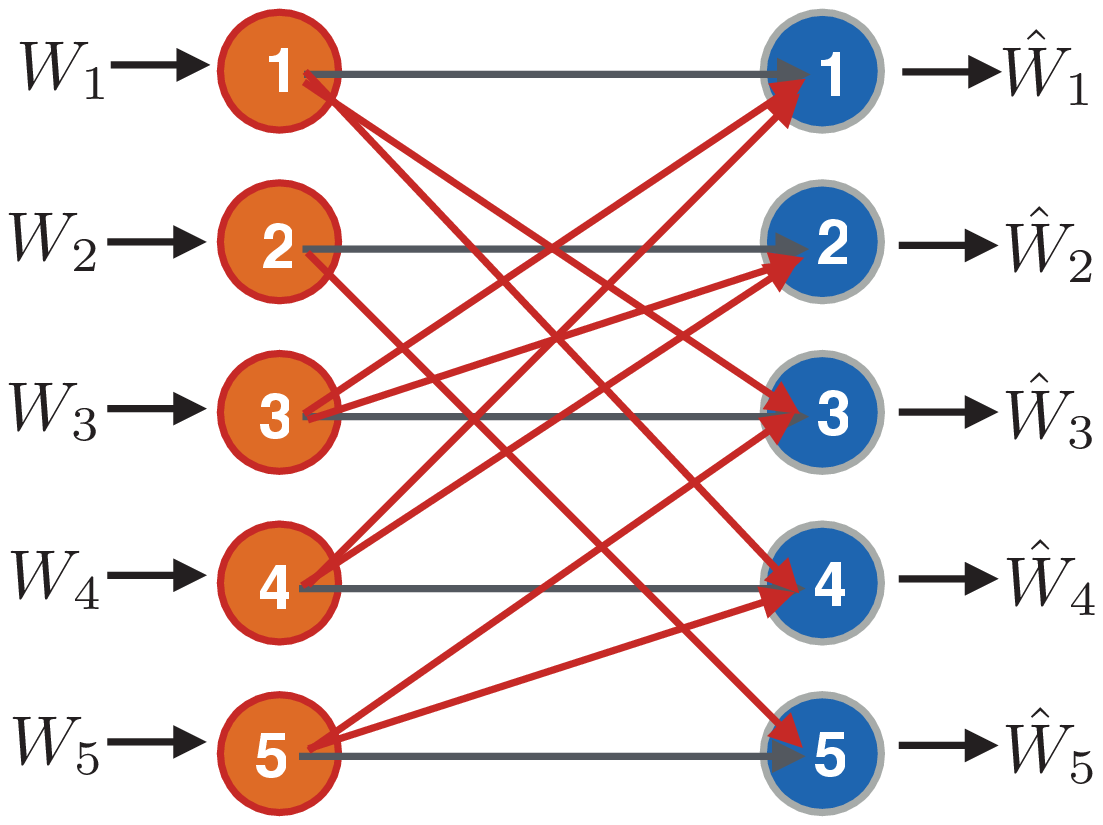}
{\tiny (a) TIM problem.}
 \end{minipage}
 \hspace*{0.4cm}
 \noindent \begin{minipage}[b]{0.31\hsize}
\centering
\includegraphics[width=\hsize]{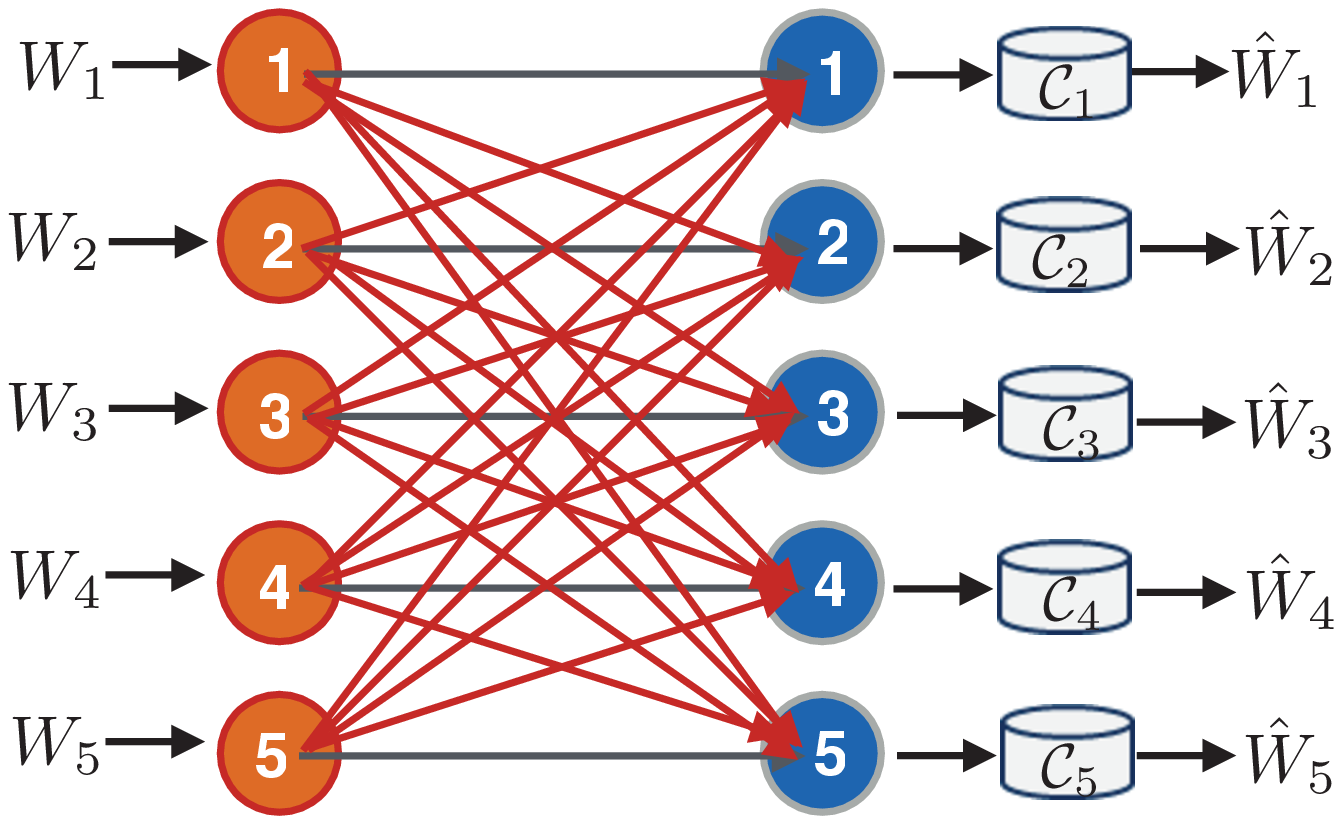}
{\tiny (b) Cache-aided interference channel.}
 \end{minipage}
 \hspace*{0.4cm}
  \noindent \begin{minipage}[b]{0.26\hsize}
\centering
\includegraphics[width=\hsize]{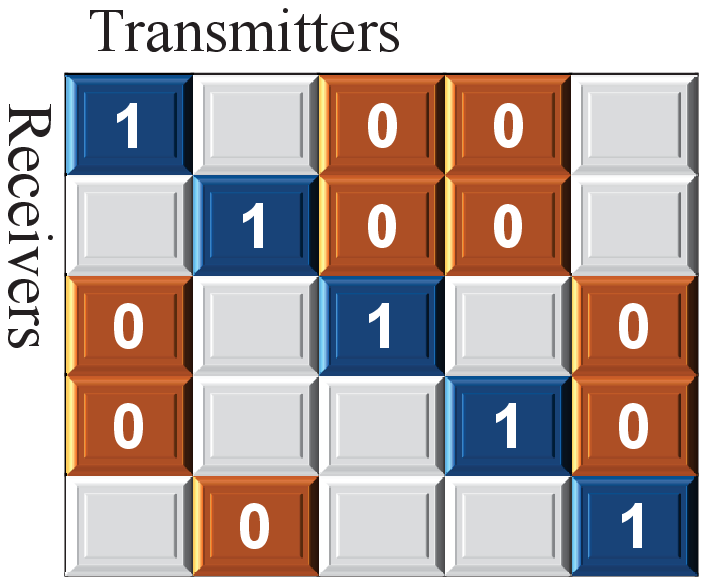}
{\tiny (c) Side information  modeling matrix.}
 \end{minipage}
 \end{tabular}
\caption{(a) A partially-connected 5-user
interference channel with the index set of connected links as  $\mathcal{V}=\{(1,
3), (1,4), (2, 3), (2, 4), (3, 1), (3, 5), (4, 1), (4, 5), (5, 1)\}$. (b) A cache-aided 5-user  interference channel with cached messages at each receiver indexed by $\mathcal{C}_1=\{2, 5\}$, $\mathcal{C}_2=\{1,5\}$, $\mathcal{C}_3=\{2,4\}$, $\mathcal{C}_4=\{2,3\}$ and $\mathcal{C}_5=\{1,3,4\}$. (c) Let $\bm{M}=[M_{ij}]\in\mathbb{C}^{K\times
K}$ with $M_{ij}=\bm{u}_i^{\sf{H}}\bm{v}_i
\in\mathbb{C}$, where $\bm{u}_i\in\mathbb{C}^n$ and $\bm{v}_i\in\mathbb{C}^{n}$ are the precoding and decoding vectors with $n$ as the number of channel uses for transmission. The incomplete matrix $\bm{M}$ with partial known entries indexed by $\mathcal{V}$ serves as the side information modeling matrix for (a) and (b), i.e., $M_{ii}=1$ means to preserve the desired signals for each receiver, $M_{ij}=0, \forall (i,j)\in\mathcal{V}$ represents cancelling the interference, and $M_{ij}=*, \forall (i,j)\notin \mathcal{V}\cup\{(i,i)\}$ can be any (unknown) values.}
\label{timfig}
\end{figure}

To exploit the full performance gains of network densification, recent years have seen progresses on interference management under various scenarios depending on the amount of shared CSI and user messages. Typical interference management strategies include interference alignment, interference coordination, coordinated multipoint transmission and reception, to name just a few. However, the significant overhead of acquiring global CSI motivates numerous research efforts on CSI overhead reduction strategies, e.g., delayed CSI, alternating CSI and mixed CSI. One of the most promising strategies is topological interference management (TIM) \cite{Jafar_TIT2013TIM}, for which only network connectivity information is required. This  is based on the fact that most of the wireless channel propagation links are weak enough to be ignored, thanks to pathloss and shadowing. However, the TIM problem turns out to be {\rev{linear}} index coding problems \cite{Jafar_TIT2013TIM}, which are in general highly intractable and only partial results exist for special cases.  Recently, a new proposal was made for the TIM problem, which can greatly assist the algorithm design. The main innovation is to model the network connectivity pattern in UDNs as an incomplete matrix. Then the TIM problem can be formulated as a generalized matrix completion problem\footnote{B. Hassibi, ``Topological interference alignment in wireless networks," {\it{Smart Antennas Workshop}}, Aug. 2014.}, which helps to develop effective linear precoding and decoding strategies. Fig. {\ref{timfig}} demonstrates the modeling framework, with Fig. {\ref{timfig}} (a) showing a 5-user interference channel as an example and Fig. {\ref{timfig}} (c) showing the corresponding modeling matrix. The task of TIM is to complete the side information modeling matrix, which will then determine the precoder and decoder \cite{Yuanming_2016LRMCTWC}.

This modeling framework is very powerful, and can be adopted to consider other design problems in UDNs. By equipping the densely deployed radio access points and mobile devices with isolated cache
storages, caching the content at the edge of the network provides a promising
way to improve
the throughput and reduce latency, as well as reducing the load of the core network and radio access networks \cite{Debbah_CMag14}. In general, content-centric communications
consist of two phases, a content placement phase followed by a content delivery
phase. However, due to the coupled wireline and wireless communications in cache-aided UDNs,
unique challenges arise in the edge caching problem. Fortunately, the incomplete matrix modeling framework can  capture the information of the content cached at different nodes. Fig. {\ref{timfig}} (b) shows an example for cache-aided 5-user interference channels, where the side information is represented in the side information modeling matrix in Fig. {\ref{timfig}} (c). Similarly, this modeling framework can also be extended to wireless distributed computing networks \cite{Ali_ComMag17}. For the prevalent distributed computing structures like MapReduce and Spark, the basic idea is that intermediate values computed in the ``map" phase based on the locally available dataset, can be regarded as the side information for the ``reduce" phase to compute the output value for a given input. This thus can help reduce the communication overhead in the ``shuffle" phase to obtain the intermediate values that are not computed locally in the ``map" phase. The incomplete matrix modeling approach will help to formulate the design problems for wireless caching and distributed computing systems.

\subsection{A Generalized Low-Rank Optimization Paradigm}

\begin{figure}[t]
\center
\includegraphics[scale = 0.6]{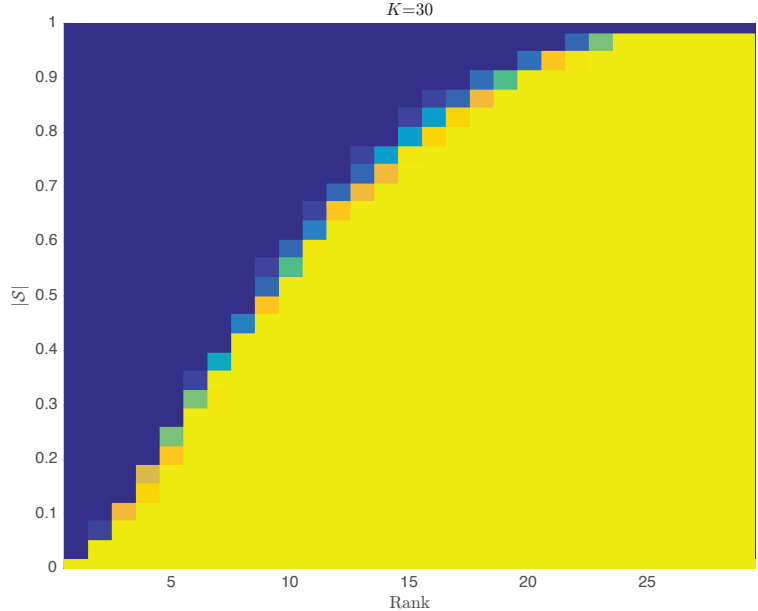}
\caption{Phase transitions for the topological interference management
problem for a partially connected $K$-user interference channel with the
network side information modeling constraint
set $\mathcal{D}=\{\bm{M}\in\mathbb{R}^{30\times 30}|M_{ii}=1,
 M_{ij}=0, \forall (i,j)\in\mathcal{S}\}$, where the set $\mathcal{S}$ is randomly
and uniformly sampled. The heat map indicates the empirical probability
of success (blue=0$\%$; yellow=100$\%$).}
\label{timnumerical}
\end{figure}

We have presented an effective and general framework to model various network side information
in UDNs. Next we present a low-rank optimization formulation to exploit the available network side information. The side information modeling matrix $\bm{M}$ as shown in Fig. {\ref{timfig}} (c) helps cancel interference over $n$ channel uses, yielding an interference-free channel with $1/n$ degrees-of-freedom (DoF), i.e., the first-order data characterization. Observe that the rank of the side information modeling matrix $\bm{M}$, denoted by ${\rm{rank}} (\bm{M})$, equals the number of channel uses $n$, which equals the inverse of the achievable DoF. To
maximize the achievable degrees-of-freedom (DoF), we thus can minimize the rank of the side information modeling matrix,  yielding the  following generalized low-rank optimization problem:  
\setlength\arraycolsep{2pt}
\begin{eqnarray}
\label{lowrank}
\mathop {\rm{minimize}}_{\bm{M}\in\mathbb{C}^{p\times q}}&&{\rm{rank}}(\bm{M})~~~~{\rm{subject~to}}~~\bm{M}\in\mathcal{D},
\end{eqnarray} 
where the constraint set $\mathcal{D}$ encodes the network side information. Low-rank optimization has been proved to be a key design tool in machine learning, high-dimensional
statistics, signal processing and computational mathematics \cite{Romberg_JSTSP16lrmc}. The rank function is nonconvex and thus is computationally difficult, but convexifying it leads to efficient algorithms. For example, the nuclear norm (i.e., the summation of singular valves of a matrix) provides a convex surrogate of the rank function that is analogous to the $\ell_1$-norm relaxation of the cardinality of a vector.

Given the special structure of the side information modeling matrix in UDNs, most existing algorithmic and theoretical results for low-rank optimization are inapplicable. The recent work \cite{Yuanming_2016LRMCTWC} contributed a novel proposal of nonconvex paradigms for solving the generalized low-rank optimization problem (\ref{lowrank})
by optimizing over the nonconvex rank constraints directly via Riemannian optimization and matrix factorization.
Fig. {\ref{timnumerical}}  illustrates the phase transition behavior for the generalized low-rank optimization in topological interference management, which characterizes the relationships between the achievable DoF and the number of connected interference links on average.  Given the rank, representing the achievable DoF, with more connected interference links, the success probability for recovering the incomplete side information modeling matrix is lower.
It thus provides the guidelines for network deployment in dense wireless networks,
content placement in cache-aided interference channels, and dataset placement in wireless distributed computing systems.

\section{Optimization Algorithms and Analysis}
\label{alg}
\begin{figure}[t]
\begin{tabular}{cc}
\hspace*{0.5cm}
\noindent \begin{minipage}[b]{0.4\hsize}
\centering
\includegraphics[width=0.85\hsize]{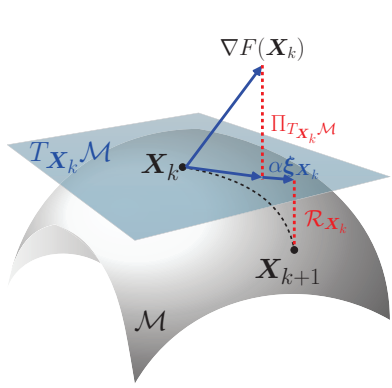}

{\scriptsize (a) Iterates of Riemannian optimization algorithm.}
 \end{minipage}
 \hspace*{0.9cm}
 \noindent \begin{minipage}[b]{0.41\hsize}
\centering
\includegraphics[width=\hsize]{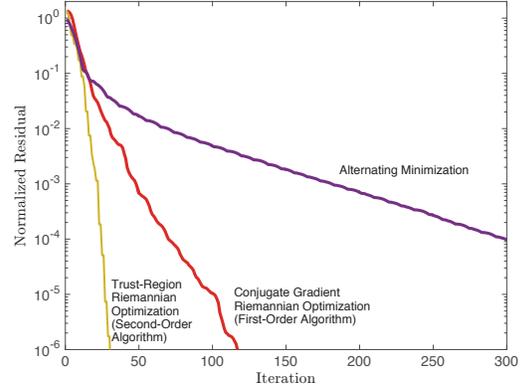}
{\scriptsize (b) Convergence rates of various algorithms.}
 \end{minipage}
 \end{tabular}
\caption{(a) Graphic representation of Riemannian optimization algorithms at
each iteration.  (b) Solving rank-constrained optimization problem $ \mathop {\rm{minimize}}_{\bm{M}\in\mathcal{M}}~f(\bm{M})$ with $\mathcal{M}=\{{\bm{M}}\in\mathbb{R}^{100\times 100}|{\rm{rank}}(\bm{M})=5\}$ and $f(\bm{M})=\sum_{i=1}^{100}(M_{ii}-1)^2+\sum_{(i,j)\in\Omega}M_{ij}^2$ ($|\Omega|=400$) via Riemannian optimization algorithms by factorizing rank-$r$ matrix $\bm{M}$, yielding a quotient manifold $[\bm{M}]=\{(\bm{U}\bm{Q}_U, \bm{Q}_U^T\bm{\Sigma}\bm{Q}_V, \bm{V}\bm{Q}_{V}): \bm{Q}_{U}, \bm{Q}_{V}\in\mathcal{Q}(r)\}$, where $\mathcal{O}(r)$ is the set of all $r\times r$ orthogonal matrices.}
\label{manopt}
\end{figure}

We have seen quite a few algorithmic challenges for the sparse and low-rank modeling frameworks for UDNs. In this section, we present some new trends in optimization  algorithms for solving the generalized sparse and low-rank optimization problems in the forms of (\ref{sparse}) and (\ref{lowrank}), respectively. Basically, numerical optimization algorithms can be classified in terms of first versus second order methods, depending on whether they use only gradient-based information versus calculations of both the first and second derivatives. The convergence rates of second-order methods are usually faster with the caveat that each iteration is more expensive. In general, there is a trade-off between the per-iteration computation cost versus the total number of iterations, though first-order methods often scale better to large-scale high-dimensional statistics problems \cite{Wainwright2014structured}. While optimization problems in communication systems are typically solved in the convex paradigm with the second-order methods, thanks to the ease of use of the CVX toolbox, we have observed the necessity of the first-order methods and the importance of the nonconvex paradigm, as will be elaborated in the following parts.

\subsection{Convex Optimization Algorithms}
We have presented a variety of methodologies to convexify the nonconvex objective functions and nonconvex constraints for the generalized sparse optimization problem (\ref{sparse}). Newton iteration based interior-point methods supported by many user-friendly software packages (e.g., CVX) provide a general way to solve constrained convex optimization problems. However, the cubic computational complexity of each Newton step limits its capability  to scale to large network sizes in UDNs. This motivates enormous research efforts to improve the computational efficiency for convex programs, including the techniques of first-order methods, randomization, parallel and distributed computing. 

Parallel and distributed optimization provides a principled way to exploit the distributed computing environments to increase the levels of scalability, while reducing the communication costs. To solve a general large-scale convex programs, a principled two-stage framework has recently been proposed in \cite{Yuanming_LargeSOCP2014} with the capability of providing certificates of infeasibility, enabling parallel and scale computing. This is achieved, in the first stage, by the matrix stuffing technique to fast transform the original convex programs into the standard conic optimization problem form via updating the associated values in the pre-stored structure of the standard conic program. In the second stage, the ADMM based algorithm is adopted to solve the standard large-scale conic optimization problem via exploiting the problem structures \cite{Yuanming_LargeSOCP2014} to enable parallel cone projection at each iterate.

Other lines of works have focused on the use of first-order methods and randomization to solve large convex programs. In particular, for sparse convex optimization problems,  Frank-Wolfe-type algorithms (a.k.a., conditional gradient) have recently gained enormous interests, fueled by the excellent scalability with projection-free operations via exploiting the well-structured sparsity constraints. The coordinate descent method has gained its popularity for scalability by choosing a single coordinate (or a block of coordinates) to be updated within each iteration, thereby reducing the iteration computing cost. Approximation techniques, including randomization methods and sketching methods, further provide algorithmic opportunities to enable scalability for, in particular, first-order methods, via speeding up numerical linear algebra or reducing problem dimensions. In particular, the stochastic gradient method provides a generic way to stochastically approximate the gradient descent method to
solve large-scale machine learning problems. All the above presented algorithmic and theoretical results may be leveraged to solve large-scale convex optimization problems in UDNs.

\subsection{Nonconvex Optimization Algorithms}
Recently, a new line of work has attracted significant attentions, which focuses on solving the nonconvex optimization problems directly via developing efficient  nonconvex procedures, sometimes with optimality guarantee. We have seen recent progress on nonconvex procedures based on various algorithms (e.g., projected/stochastic/conditional gradient methods, Riemannian manifold optimization algorithms) for a class of high-dimensional statistical problems and machine learning problems, including low-rank matrix completion, phase retrieval and blind deconvolution, to name just a few. In particular, optimization by directly exploiting problems' manifold structures is becoming a general and powerful approach to solve various nonconvex optimization problems. The structured constraints such as rank and orthogonality appear in many machine learning applications, including sensor network localization, dimensionality reduction, low-rank matrix recovery, phase synchronization, and community detection.   

At a high-level standpoint, Riemannian optimization is the extension of standard unconstrained optimization  searching in the Euclidean space to optimizing in the Riemannian manifold space  by generalizing the concepts such as the gradient and Hessian \cite{manopt}. A graphic representation of  Riemannian optimization algorithms is illustrated in Fig. {\ref{manopt}}. Specifically, the Euclidean gradient $\nabla f(\bm{X}_k)$ needs to be projected to
the tangent space $T_{\bm{X}_k}\mathcal{M}$ of manifold $\mathcal{M}$ to
define  a search direction $\bm{\xi}_{\bm{X}_k}$ (which can be computed based on the principles of the conjugate gradient method or trust-region method), followed
by the retraction operator $\mathcal{R}_{\bm{X}_k}$ to define a new iterate
$\bm{X}_{k+1}=\mathcal{R}_{{\bm{X}}_k}(\alpha\bm{\xi}_k)$ ($\alpha$ is the step size) on the manifold $\mathcal{M}$.
In particular, we exploit the manifold geometry of fixed-rank matrices to solve the low-rank optimization problem (\ref{lowrank}) efficiently. Fig. {\ref{manopt}} (b) demonstrates the effectiveness of Riemannian optimization based methods. It shows that the Riemannian optimization enjoys fast convergence rates, e.g., compared with an existing approach based on alternating minimization.

\section{{\rev{Conclusions}} and Future Directions}
This article presented generalized sparse and low-rank optimization techniques for optimizing across communication, computation and storage resources in UDNs by exploiting  network structures and side information. Illustrated by important application examples, various structured sparse modeling methods were introduced, and an incomplete matrix representation was presented to model different types of network side information. Methodologies of designing scalable algorithms were discussed, including both convex and nonconvex methods.
The presented results and methodologies demonstrated the effectiveness of structured optimization techniques for designing UDNs.

Despite the encouraging progress, there still remain a variety of interesting open questions. To date, generalized sparse and low-rank optimization techniques are mainly applied to improve the network energy efficiency and spectral efficiency in UDNs. However, emerging mobile applications have strong demands for user privacy and ultra-low latency communications, which call for more general mathematical models and formulations. Other interesting problems concern the theoretical analysis for the generalized sparse and low-rank optimization models and algorithms. Although we have seen significant progresses for theoretical understanding of sparse and low-rank optimization problems via convex relaxation approaches \cite{Tropp_livingedge2014} and nonconvex procedures, it is challenging to apply existing results to the generalized sparse and low-rank optimization problems (\ref{sparse}) and (\ref{lowrank})  due to the complicated structures. Finally, there are a variety of interesting research directions associated with improving the computational scaling behaviour
of various algorithms via recent proposals, e.g., randomized algorithms based on sketching.

\bibliographystyle{ieeetr}

\begin{thebibliography}{10}

\bibitem{Tony_CRAN17}
T.~Q.~S. Quek, M.~Peng, O.~Simeone, and W.~Yu, {\em Cloud Radio Access
  Networks: Principles, Technologies, and Applications}.
\newblock Cambridge University Press, 2017.

\bibitem{Haijun_WCmag16}
H.~Zhang, Y.~Dong, J.~Cheng, M.~J. Hossain, and V.~C.~M. Leung, ``Fronthauling
  for 5{G} {LTE-U} ultra dense cloud small cell networks,'' {\em IEEE Wireless
  Commun. Mag.}, vol.~23, pp.~48--53, Dec. 2016.

\bibitem{Ververidis_CST14}
G.~Xylomenos, C.~N. Ververidis, V.~A. Siris, N.~Fotiou, C.~Tsilopoulos,
  X.~Vasilakos, K.~V. Katsaros, and G.~C. Polyzos, ``A survey of
  information-centric networking research,'' {\em IEEE Commun. Surveys Tuts.},
  vol.~16, pp.~1024--1049, Second 2014.

\bibitem{Debbah_CMag14}
E.~Bastug, M.~Bennis, and M.~Debbah, ``Living on the edge: The role of
  proactive caching in 5g wireless networks,'' {\em IEEE Commun. Mag.},
  vol.~52, pp.~82--89, Aug. 2014.

\bibitem{Yuanming_LargeSOCP2014}
Y.~Shi, J.~Zhang, B.~O'Donoghue, and K.~Letaief, ``Large-scale convex
  optimization for dense wireless cooperative networks,'' {\em IEEE Trans.
  Signal Process.}, vol.~63, pp.~4729--4743, Sept. 2015.

\bibitem{manopt}
N.~Boumal, B.~Mishra, P.-A. Absil, and R.~Sepulchre, ``{M}anopt, a {M}atlab
  toolbox for optimization on manifolds,'' {\em J. Mach. Learn. Res.}, vol.~15,
  pp.~1455--1459, 2014.

\bibitem{Yuanming_TWC2014}
Y.~Shi, J.~Zhang, and K.~B. Letaief, ``Group sparse beamforming for green
  {C}loud-{RAN},'' {\em IEEE Trans. Wireless Commun.}, vol.~13, pp.~2809--2823,
  May 2014.

\bibitem{Yuanming_2016LRMCTWC}
Y.~Shi, J.~Zhang, and K.~B. Letaief, ``Low-rank matrix completion for
  topological interference management by {R}iemannian pursuit,'' {\em IEEE
  Trans. Wireless Commun.}, vol.~15, pp.~4703--4717, Jul. 2016.

\bibitem{Thomas_IEEEAccess20155G}
G.~Wunder, H.~Boche, T.~Strohmer, and P.~Jung, ``Sparse signal processing
  concepts for efficient 5{G} system design,'' {\em IEEE Access}, vol.~3,
  pp.~195--208, 2015.

\bibitem{Haijun_JSAC17}
H.~Zhang, S.~Huang, C.~Jiang, K.~Long, V.~C.~M. Leung, and H.~V. Poor, ``Energy
  efficient user association and power allocation in millimeter-wave-based
  ultra dense networks with energy harvesting base stations,'' {\em IEEE J.
  Sel. Areas Commun.}, vol.~35, pp.~1936--1947, Sept. 2017.

\bibitem{Jafar_TIT2013TIM}
S.~Jafar, ``Topological interference management through index coding,'' {\em
  IEEE Trans. Inf. Theory}, vol.~60, pp.~529--568, Jan. 2014.

\bibitem{Ali_ComMag17}
S.~Li, M.~A. Maddah-Ali, and A.~S. Avestimehr, ``Coding for distributed fog
  computing,'' {\em IEEE Commun. Mag.}, vol.~55, pp.~34--40, Apr. 2017.

\bibitem{Wainwright2014structured}
M.~J. Wainwright, ``Structured regularizers for high-dimensional problems:
  Statistical and computational issues,'' {\em Annu. Rev. Stat. Appl.}, vol.~1,
  pp.~233--253, 2014.

\bibitem{Romberg_JSTSP16lrmc}
M.~A. Davenport and J.~Romberg, ``An overview of low-rank matrix recovery from
  incomplete observations,'' {\em IEEE J. Sel. Topics Signal Process.},
  vol.~10, pp.~608--622, Jun. 2016.

\bibitem{Tropp_livingedge2014}
D.~Amelunxen, M.~Lotz, M.~B. McCoy, and J.~A. Tropp, ``Living on the edge:
  phase transitions in convex programs with random data,'' {\em Inf.
  Inference}, vol.~3, pp.~224--294, Jun. 2014.

\end{thebibliography}
%
%

\section*{Biographies}

\begin{description}
\item[Yuanming Shi] [S'13-M'15] (shiym@shanghaitech.edu.cn) received the B.S. degree from Tsinghua University in 2011, and the Ph.D. degree from The Hong Kong University of Science and Technology (HKUST) in 2015. He is currently an Assistant Professor at ShanghaiTech University. He received the 2016 IEEE Marconi Prize Paper Award and the 2016 Young Author Best Paper Award by the IEEE Signal Processing Society. His research interests include dense wireless networks, intelligent IoT, mobile AI,  machine learning, statistics, and optimization.

\item[Jun Zhang] [M'10-SM'15] (eejzhang@ust.hk) received the Ph.D. degree from the University of Texas at Austin. He is currently a Research Assistant Professor at Hong Kong University of Science and Technology. He received the 2016 Marconi Prize Paper Award in Wireless Communications, and the 2016 IEEE ComSoc Asia-Pacific Best Young Researcher Award. His research interests include dense wireless cooperative networks, mobile edge caching and computing, cloud computing, and big data analytics systems.

\item[Wei Chen] [S'05-M'07-SM'13] (wchen@tsinghua.edu.cn) received his BS and Ph.D. degrees (Hons.) from Tsinghua University in 2002 and 2007, respectively. Since 2007, he has been on the faculty at Tsinghua University, where he is a tenured full Professor and a member of the University Council. He is a member of the National 10000-Talent Program and a Cheung Kong Young Scholar. He received the IEEE Marconi Prize Paper Award and the IEEE Comsoc Asia Pacific Board Best Young Researcher Award.

\item[Khaled B. Letaief] [S''85-M''86-SM''97-F''03] (eekhaled@ust.hk) received Ph.D. Degree from Purdue University, USA. From 1990 to 1993, he was faculty member at University of Melbourne, Australia.  He has been with HKUST since 1993 where he was Dean of Engineering.  From September 2015, he joined HBKU in Qatar as Provost.  He is Fellow of IEEE, ISI Highly Cited Researcher, and recipients of many distinguished awards.  He served in many IEEE leadership positions including ComSoc Vice-President for Technical Activities and Vice-President for Conferences.
\end{description}

\end{document}